\documentclass[conference]{IEEEtran}
\IEEEoverridecommandlockouts
\usepackage[margin=1in]{geometry}
\usepackage[T1]{fontenc}
\usepackage{amsfonts}
\usepackage{multirow}
\usepackage{dsfont}
\usepackage{bbm}
\usepackage{booktabs}
\usepackage{xcolor}
\usepackage{enumitem}  % For enhanced itemize environments with options like [leftmargin=*]
\usepackage{pifont}    % For \ding{} command to access circled numbers and other symbols
\usepackage{amsmath}
\usepackage{graphicx}
\usepackage{cite}
\usepackage{amsmath,amssymb,amsfonts}
\usepackage{algorithmic}
\usepackage{graphicx}
\usepackage{textcomp}
\usepackage{xcolor}
\usepackage{multirow}
\def\BibTeX{{\rm B\kern-.05em{\sc i\kern-.025em b}\kern-.08em
    T\kern-.1667em\lower.7ex\hbox{E}\kern-.125emX}}
\begin{document}

\title{Mind the Context: Attention-Guided Weak-to-Strong Consistency for Enhanced Semi-Supervised Medical Image Segmentation
}
%\title{Conference Paper Title*\\
%{\footnotesize \textsuperscript{*}Note: Sub-titles are not captured for https://ieeexplore.ieee.org  and

\author{
\IEEEauthorblockN{Yuxuan Cheng$^\dagger$}
\IEEEauthorblockA{
\textit{College of Engineering} \\
\textit{Huazhong Agricultural University} \\
Wuhan, China \\
hanxuanwxss@gmail.com}
\and
\IEEEauthorblockN{Chenxi Shao$^\dagger$}
\IEEEauthorblockA{
\textit{College of Informatics} \\
\textit{Huazhong Agricultural University} \\
Wuhan, China \\
alicespring@webmail.hzau.edu.cn}
\and
\IEEEauthorblockN{Jie Ma}
\IEEEauthorblockA{
\textit{College of Informatics} \\
\textit{Huazhong Agricultural University} \\
Wuhan, China \\
margo820@webmail.hzau.edu.cn}
\and
\IEEEauthorblockN{Yunfei Xie}
\IEEEauthorblockA{
\textit{School of Automation and Artificial Intelligence} \\
\textit{Huazhong University of Science and Technology} \\
Wuhan, China \\
xieyunfei01@gmail.com}
\and
\IEEEauthorblockN{Guoliang Li$^*$}
\IEEEauthorblockA{
\textit{College of Informatics} \\
\textit{Huazhong Agricultural University} \\
Wuhan, China \\
guoliang.li@mail.hzau.edu.cn}
}

\maketitle
\begingroup
\renewcommand\thefootnote{}\footnotetext{$^\dagger$ Equal contribution. $^*$ Corresponding author.}
\endgroup
\begin{abstract}
Medical image segmentation is a pivotal step in diagnostic and therapeutic processes, relying on high-quality annotated data that is often challenging and costly to obtain. Semi-supervised learning offers a promising approach to enhance model performance by leveraging unlabeled data. Although weak-to-strong consistency is a prevalent method in semi-supervised image segmentation, there is a scarcity of research on perturbation strategies specifically tailored for semi-supervised medical image segmentation tasks. To address this challenge, this paper introduces a simple yet efficient semi-supervised learning framework named Attention-Guided weak-to-strong Consistency Match (AIGCMatch). The AIGCMatch framework incorporates attention-guided perturbation strategies at both the image and feature levels to achieve weak-to-strong consistency regularization. This method not only preserves the structural information of medical images but also enhances the model's ability to process complex semantic information. Extensive experiments conducted on the ACDC dataset have validated the effectiveness of AIGCMatch. Our method achieved a 90.4\% Dice score in the 7-case scenario on the ACDC dataset, surpassing the state-of-the-art methods and demonstrating its potential and efficacy in clinical settings. 
%Additionally, on the ISIC-2017 dataset, we significantly outperformed our baseline, indicating the robustness and generalizability of AIGCMatch across different medical image segmentation tasks.
\end{abstract}


\section{Introduction}
\label{sec:intro}
%% revised by zhy
Medical image segmentation is a fundamental task aimed at delineating anatomical structures such as organs and tumors by classifying each pixel into distinct categories \cite{van2011computer,litjens2017survey}. The precision achieved through medical image segmentation provides invaluable volumetric and shape details, serving as foundational elements for disease diagnosis, treatment efficacy evaluation, and quantitative analysis within clinical settings \cite{oreiller2022head,lalande2022deep}. Data-driven approaches, particularly those based on fully supervised learning, have shown remarkable performance in this domain \cite{ronneberger2015u,cciccek20163d,zhou2018unet++}. However, their success relies heavily on the availability of large-scale, high-quality annotated datasets, which are often challenging and expensive to obtain \cite{litjens2017survey}. This limitation has spurred the development of semi-supervised learning (SSL) approaches, which leverage abundant unannotated data alongside a smaller set of annotated examples \cite{oliver2018realistic,zhang2023multi}.

\begin{figure*}[t]
\centering
\label{fig:cat}
\includegraphics[width=0.8\textwidth, keepaspectratio]{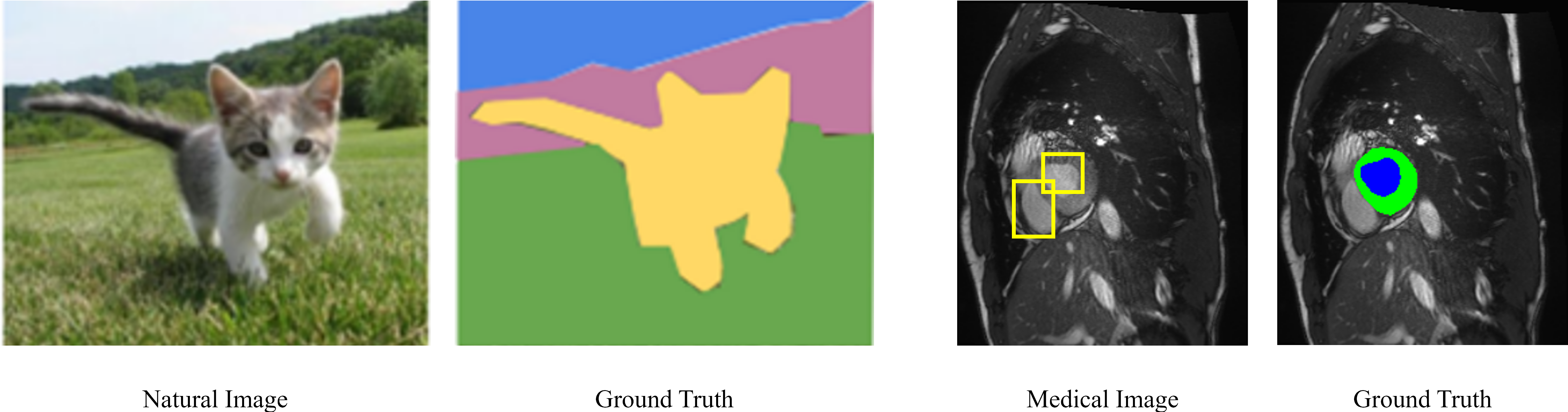}
\setlength{\belowcaptionskip}{-0.78cm} %调整图片标题与下文距离
\caption{An example of the difference between a natural image and medical image. 
In natural images, there is a clear pixel-level difference between the foreground, such as a cat, and the background, such as grassland. In contrast, within medical imaging, the two parts enclosed by the yellow boxes do not exhibit significant pixel-level differences; however, they represent semantically distinct entities.}
\end{figure*}

In the domain of SSL, consistency regularization has emerged as a key technique to use unannotated data to improve model performance \cite{oliver2018realistic,bortsova2019semi,li2020transformation,luo2021semi,wang2020double,zhu2021improving}. The principle behind consistency regularization requires the model to produce consistent predictions on different perturbations of the input data. This is particularly relevant in medical imaging, where slight variations in image acquisition parameters or patient positioning can lead to variations in the appearance of the same anatomical structures \cite{litjens2017survey}. However, despite the remarkable performance of consistency regularization in general computer vision tasks, its direct application to medical image segmentation faces two significant challenges that limit its effectiveness in clinical applications:

\textbf{C1. How can we design perturbation strategies that respect the unique characteristics of medical images?} Existing approaches predominantly rely on random perturbation techniques originally designed for natural images, which fail to account for the complex anatomical structures present in medical data \cite{Yang_2023_CVPR}. As illustrated in Fig.~\ref{fig:cat}, natural images typically exhibit distinct pixel-level differences between objects and backgrounds, making them amenable to random perturbation techniques. For instance, a cat can be easily differentiated from its background due to clear visual boundaries and color contrasts. In contrast, medical images contain complex anatomical structures with subtle intensity variations between different tissues, where adjacent regions may appear visually similar yet represent entirely different anatomical entities. These subtle differences require preservation of specific structural and contextual information that random perturbations may inadvertently disrupt. Consequently, generic perturbation strategies may compromise the semantic integrity essential for accurate medical image segmentation, highlighting the need for perturbation approaches specifically tailored to the unique characteristics of medical imaging data.

\textbf{C2. How can we extend the scope of perturbations beyond the image level to capture high-dimensional semantic information?} Current methods predominantly focus on perturbations at the image level, overlooking the potential of feature-level perturbations to enhance model robustness. Medical images contain complex high-dimensional semantic information that is progressively extracted and refined through different layers of deep learning models. By restricting perturbations to the input level, these methods fail to challenge the model's ability to maintain consistent representations across different feature abstraction levels. This limitation is particularly significant in medical image segmentation, where subtle feature variations can substantially impact segmentation boundaries. A comprehensive approach that incorporates perturbations at multiple levels of representation could potentially improve the model's capacity to capture and maintain consistent semantic information throughout the segmentation process.

To address these challenges, we propose \textbf{AttentIon-Guided Consistency regularization Match (AIGCMatch)}, a novel semi-supervised framework for medical image segmentation. \ding{182} To address the first challenge of inappropriate perturbation strategies, we introduce attention-guided CutMix (AttCutMix), which leverages the model's attention mechanisms to identify semantically meaningful regions for targeted perturbations. Unlike random perturbations such as CutMix \cite{yun2019cutmix}, AttCutMix preserves the structural integrity of medical images while introducing beneficial variability, ensuring that perturbations respect anatomical boundaries and tissue relationships. \ding{183} To address the second challenge of limited perturbation scope, we propose attention-guided feature perturbation (AttFeaPerb), a novel technique that extends consistency regularization to the feature level by selectively perturbing channels with varying attention scores. This approach establishes a weak-to-strong consistency paradigm at the feature level, enhancing the model's ability to maintain robust representations across different levels of abstraction. By integrating these attention-guided perturbation strategies at both image and feature levels, AIGCMatch effectively captures the complex structural and high-dimensional semantic information in medical images, significantly improving segmentation performance while reducing the requirement for labeled data.

Our contributions are summarized as follows:
\begin{itemize}[leftmargin=*]
    \item We present a simple yet efficient semi-supervised framework named \textbf{AIGCMatch} for the practical medical image segmentation problem by taking advantage of a large amount of unlabeled data, which largely reduces annotation efforts for cardiac surgeons.
    
    \item We explore an effective perturbation strategy that uses the model's own attention to guide perturbations at image and feature levels.
    
    \item Through extensive experimentation on the ACDC dataset, we validate the efficacy of \textbf{AIGCMatch}. Our framework consistently outperforms state-of-the-art methods, demonstrating its effectiveness and potential for widespread adoption in clinical settings.
\end{itemize}

\section{Related Work}
\subsection{Medical Image Analysis}
Medical image analysis is crucial for computer-aided diagnosis and treatment planning, with applications in identifying pathological tissues and tracking disease progression~\cite{jiao2023learning}. Advances in imaging modalities like CT, MRI, and US have significantly enhanced the visualization of internal structures. A pivotal task in this field is medical image segmentation, which aids in precise diagnoses and planning by delineating regions of interest.
The advent of deep learning, particularly Encoder-Decoder architectures, has revolutionized medical image segmentation. Early network design breakthroughs included the U-Net~\cite{ronneberger2015u}, which introduced a symmetric encoder-decoder structure and skip connections, setting new standards for accuracy and efficiency.
Building on U-Net's success, variants like U-Net++~\cite{zhou2018unet++} and U-Net3+~\cite{huang2020unet} were proposed to improve feature fusion and segmentation performance with additional enhancements. The emergence of the Visual Transformer (ViT) model has led to hybrid architectures combining CNNs for local spatial information and Transformers for long-range dependencies and semantic processing~\cite{zhang2021transfuse,manzari2024befunet}.

\subsection{Semi-supervised medical image segmentation}

%\subsection{Semi-supervised medical image segmentation}
However, medical image segmentation based fully supervised are heavily dependent on large volumes of expert-annotated data, which is a time-consuming, labor-intensive, and expensive process \cite{leiner2019machine}. 
%Additionally, these models may suffer from reduced effectiveness when encountering novel data or shifts in data distribution, thereby constraining their generalizability~\cite{li2020transformation}.
To address the question, researchers have introduced semi-supervised learning into medical image segmentation, and they primarily include pseudo-labeling,  generative models, and autoencoders.

\textbf{Pseudo-labeling} involves utilizing predictions from a model trained on labeled data to generate pseudo-labels for unlabeled data. These pseudo-labels are then used to augment the training set and train the model iteratively~\cite{zhang2022boostmis,10250843,lu2023uncertainty}.

%\textbf{Graphical model} approaches to leverage the structure of graphs to represent pixel relationships within images, maximizing the posterior probability of labeled data for segmentation\cite{shakeri2016sub,alansary2016fast,cai2016pancreas}. As highlighted by Litjens et al.\cite{litjens2017survey}, these methods typically construct an energy function that models the relationships between pixel labels and neighboring pixels, promoting label consistency within similar regions. Although these approaches perform well on small-scale datasets, they often face computational complexity and a lack of labeled information when applied to large-scale imaging data.

\textbf{Generative models}, particularly Generative Adversarial Networks (GANs), have also gained widespread application in semi-supervised medical image segmentation\cite{souly2017semi,laine2016temporal,hung2018adversarial}. Ouali et al.\cite{ouali2020semi} note that GANs, through the adversarial training of generators and discriminators, not only generate realistic samples but also effectively leverage unlabeled data. This capability enables models to achieve improved segmentation performance with limited labeled samples.

\textbf{Autoencoder} is another commonly used method in semi-supervised learning\cite{chen2019multi,sajjadi2016regularization}, which compresses the input images into lower-dimensional representations and then reconstructs the original images. As mentioned by Sajjadi et al. \cite{sajjadi2016regularization}, through self-supervised learning on unlabeled data, autoencoders can extract important features that provide richer information for segmentation tasks.

\textbf{Consistency Regularization}
While these previous methods laid the foundation for subsequent techniques, they still encounter challenges in dealing with imbalanced data and noise in complex scenarios.

Consistency Regularization
\cite{bortsova2019semi,basak2022exceedingly,luo2021semi} is the most popular technique in semi-supervised medical image segmentation. Many works focus on ensuring the model outputs the same result under different perturbations. Basak et al. propose a method of consistency prediction using interpolated unlabeled images to improve the performance of semi-supervised medical image segmentation\cite{basak2022exceedingly}. Luo et al. proposed a semi-supervised medical image segmentation framework leveraging dual-task consistency to improve segmentation accuracy by jointly predicting pixel-wise segmentation and geometry-aware level representations\cite{luo2021semi}.

\begin{figure*}[t]
\centering
\includegraphics[width=.9\linewidth, keepaspectratio]{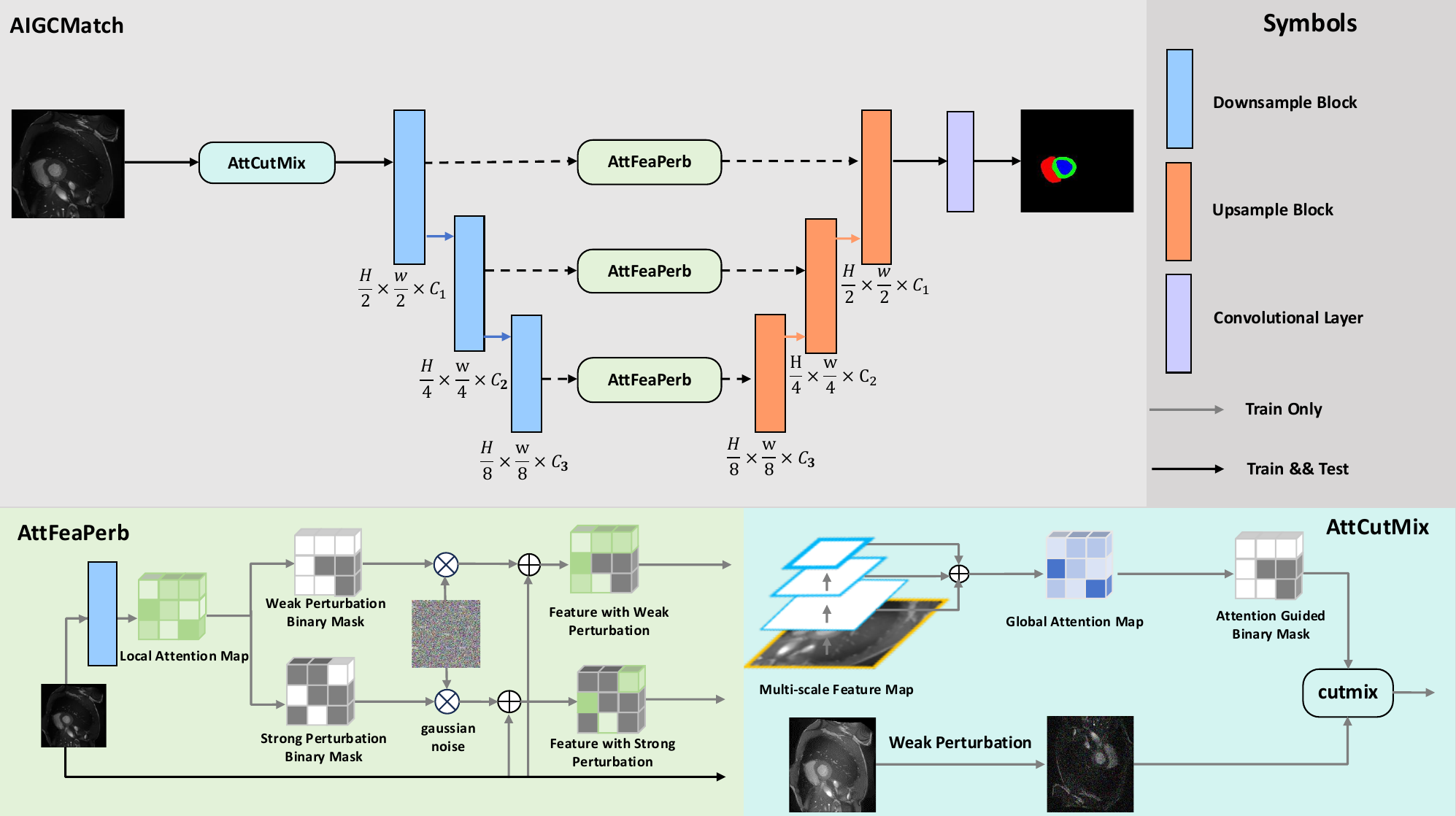}
 \caption{Our proposed method, AIGCMatch. Labeled data ignores the perturbation for supervised training. As for unlabeled data, we use AttCutMix for image perturbation and AttFeaPerb for feature perturbation to self-supervised training.}
 \label{fig:framework}
\end{figure*}
\section{Method}
\subsection{The Overview of Framework}
\begin{figure}
\centering
\begin{tabular}{cc}
\includegraphics[width=3.33cm]{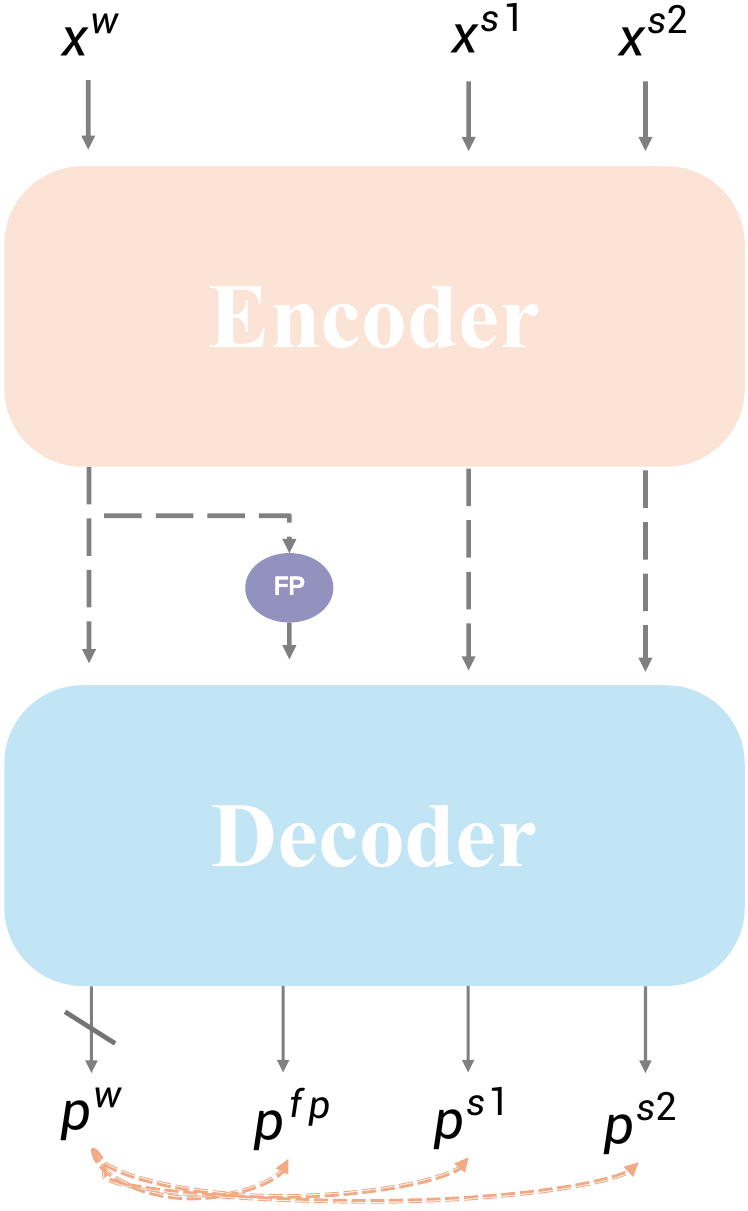}&
\includegraphics[width=3.33cm]{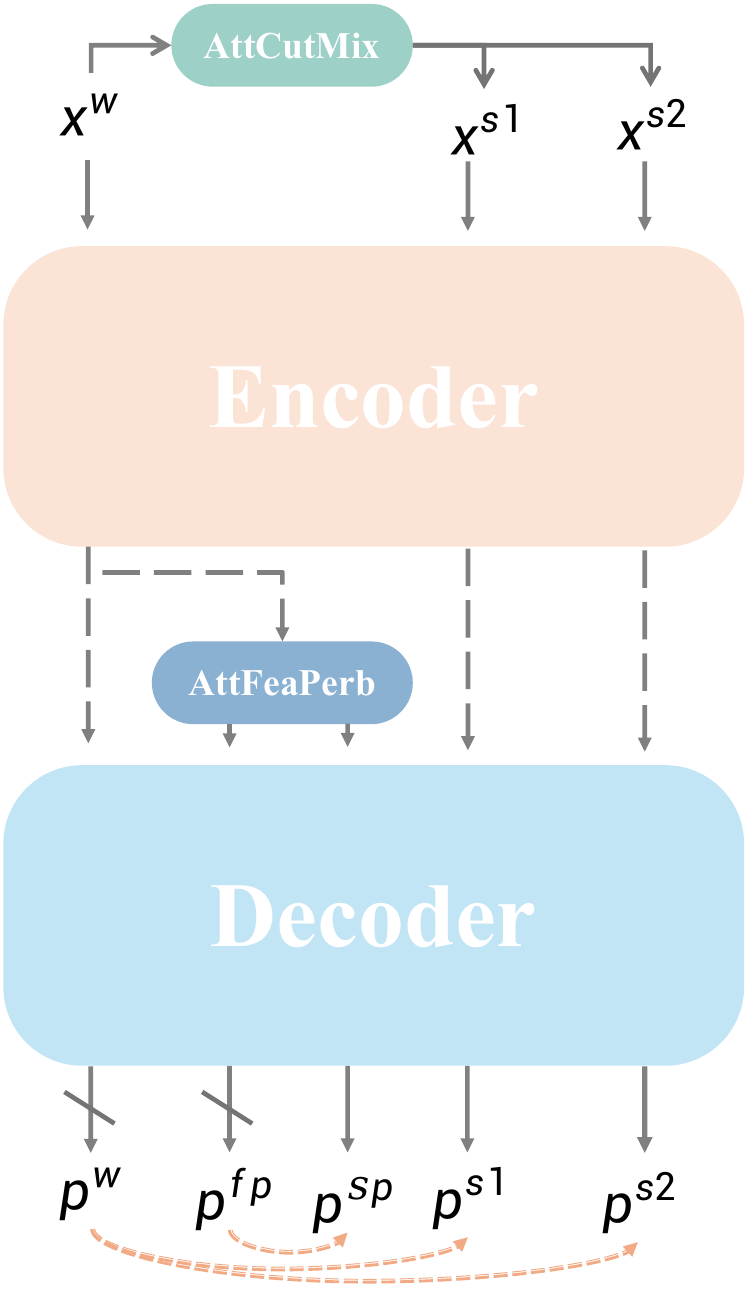}\\
(a) UniMatch & (b) AIGCMatch
\end{tabular}
\caption{Comparison with UniMatch and AIGCMatch. (a)The UniMatch baseline. The Fp denotes feature perturbation. (b) Our proposed AIGCMatch. AttCutMix denotes our method of image perturbation, while AttFeaPerb denotes our method of feature perturbation, and the \textcolor{orange}{dashed curves} represent supervision.}
\label{fig:compar}
\vspace{-0.7cm}
\end{figure}

As depicted in Fig~\ref{fig:framework}, the proposed framework consists of two attention-guided perturbation modules in the image and feature level, aiming to enhance the capability of the semi-supervised segmentation model to capture complex structural and high-dimensional semantic information in medical images. To enhance the pertinence of perturbations to the model, we fully leverage the attention from various parts of the model. At each level, the perturbations are crafted according to the attention derived from the respective parts of the model.
In \textcolor{red}{Fig~\ref{fig:compar}}, we compared the differences between our model and the Unimatch\cite{Yang_2023_CVPR}. It comprises a supervised branch and two unsupervised branches. The model is supervised for labeled data by computing the loss between the model's predictions and the ground truth (GT). For unlabeled data, attention maps from the model are employed to guide perturbations at both the feature and the image levels. Specifically, at the feature level, we proposed AttFeaPerb to obtain strong perturbation predictions $p^{fsp}\in \mathbb{R}^{C\times H\times W}$ and weak perturbation predictions $p^{fwp}\in \mathbb{R}^{C\times H\times W}$. At the image level, we used predictions $p^{w}\in \mathbb{R}^{C\times H\times W}$ obtained from weak perturbations to supervise the predictions $p^{s1}\in \mathbb{R}^{C\times H\times W}$ and $p^{s2}\in \mathbb{R}^{C\times H\times W}$ obtained from AttCutMix.

\subsection{Attention-Guided Perturbation in Image Level}
In previous studies, randomization methods such as CutMix have been the most common approach to introduce strong perturbations at the image level in semi-supervised image segmentation. However, these methods do not consider the actual semantics of the cropped content \cite{liu2022tokenmix}, which can disrupt contextual and semantic information within the image and weaken the model's ability to extract deep-level image features. 

Building upon the foundation of CutMix for the implementation of strong perturbations at the image level, we introduce a variant termed Attention-guided CutMix (AttCutMix). This method leverages the model's attention mechanism to identify and select regions within the training image that exhibit heightened attention scores. These regions are then strategically swapped with corresponding segments from other images, thereby synthesizing novel training data that can be utilized for unsupervised training phases within our model's regimen. 

To describe the Attention-guided CutMix method (AttCutMix), let $x\in \mathbb{R}^{C\times H\times W}$ represent the input image, and let $y\in \mathbb{R}^{H\times W}$ represent the model's prediction for the input image. $(x_A, y_A)$and $(x_B, y_B)$ represent two different sets of unlabeled training data together with the predictions of the model for these inputs.$(\hat{x}, \hat{y})$ represents the new training data synthesized by the AttCutMix method. The operation of AttCutMix can be formalized through the following set of equations:
\begin{equation}
M = A_{decoder} > \gamma
\end{equation}
\begin{equation}
\hat{x} = M \odot x_A + ( 1 - M ) \odot x_B
\end{equation}
\begin{equation}
\hat{y} = M \odot y_A + ( 1 - M ) \odot y_B
\end{equation}
where $A_{decoder}$ refers to the attention map output by the model's decoder component, $\gamma$ denotes a pre-defined hyperparameter that is instrumental in controlling the generation of a binary matrix. Specifically, $M$ is a binary matrix denoted as $M \in \{0,1\}^{H\times W}$, which means the regions of the image that are designated for exchange. By fine-tuning the hyperparameter, the objective of AttCutMix is to ensure the preservation of semantic information in the salient parts of the image, while simultaneously introducing beneficial perturbations. In addition to this, we have adopted the Dual-stream Perturbation approach from Unimatch, utilizing AttCutMix to generate two distinct strongly perturbed images for unsupervised training.
\subsection{Exploring Weak-to-Strong Consistency in the Feature Level}
In previous studies, discussions and definitions regarding the consistency of weak and strong perturbations have been confined to disturbances at the image level. Disturbances such as image rotation, translation, and scaling have been categorized as weak perturbations, while color transformations and CutMix operations have been defined as strong perturbations. However, no clear distinction has been made between weak and strong perturbations at the feature level.

Inspired by the concept of weak and strong consistency at the image level, we propose the notion of weak-to-strong consistency regularization at the feature level. In the context of feature-level perturbations, we define the addition of a perturbation to channels with lower attention within the model as a weak perturbation operation. Conversely, adding perturbation to channels with higher attention within the model is a strong perturbation operation. Empirical evidence suggests that the quality of the model's predictive outcomes for features subjected to weak perturbations is superior to those for features experiencing strong perturbations. Consequently, during the model training process, we utilize the model's predictions on weakly perturbed features to supervise the predictions of strongly perturbed features, which we call attention-guided feature perturbation (AttFeaPerb).

For perturbations at the feature level, we introduce noise to the feature maps corresponding to each downsampling module of the model encoder based on the attention given to the image by that module. These perturbed feature maps are then conveyed to the upsampling modules of the corresponding decoder through the skip connection architecture of the U-Net. To reduce the learning difficulty for the model, we gradually increase the intensity of the feature-level perturbation by the number of training epochs. 

We define $F_i\in \mathbb{R}^{C_i×H_i×W_i }$as the feature map of the $i$-th downsampling module in the encoder. $F_i^w\in \mathbb{R}^{C_i×H_i×W_i }$ represents the characteristic vector of the $i$-th downsampling module after applying weak perturbation. $F_i^S\in \mathbb{R}^{C_i×H_i×W_i }$denotes the feature vector of the $i$-th downsampling module following the application of strong perturbation. $A_i\in \mathbb{R}^{H_i×W_i }$ signifies the heatmap of the $i$-th downsampling module. The operations for perturbation at the feature level can be expressed with the following formulas:
\begin{equation}
M_i = A_i > \beta
\end{equation}
\begin{equation}
F_i^W=F_i+G(F_i )\odot(t(x)\times M_i)
\end{equation}
\begin{equation}
F_i^S=F_i+G(F_i )\odot (t(x)\times (1-M_i)+\phi)
\end{equation}
where $\beta$ is a pre-defined hyperparameter utilized for controlling the generation of matrix $M_i$, The matrix $M_i \in \{0,1\}^{H\times W}$is a binary matrix indicating which locations in the feature map should be perturbed. $G(F_i)$ is a Gaussian noise generation function designed to produce a Gaussian noise matrix of the same shape as $F_i$. $t(x)$ is a monotonically increasing function that controls the intensity of the perturbation, and $\phi$ is a bias term. Specifically, the function $t(x)$ can be represented in the following form:
\begin{equation}
    t(x)=1-k\times (1-\frac{x}{epoch})^{0.9}
\end{equation}
In the above equation, $k$ is a pre-defined hyperparameter that represents the maximum intensity of the perturbation to be added. $x$ denotes the current epoch number during training, and $epoch$ signifies the total number of training epochs.
\subsection{Loss Function}
In semi-supervised image segmentation tasks, the loss function typically consists of two components: supervised loss and unsupervised loss. We denote the loss function as $\mathcal{L}$, the supervised loss as $\mathcal{L}_x$, and the unsupervised loss as $\mathcal{L}_u$. The final loss function can be formally expressed as:
\begin{equation}
\mathcal{L}=\frac{1}{2}(\mathcal{L}_x+\mathcal{L}_u)
\end{equation}
For the supervised loss, we employ both cross-entropy loss and Dice coefficient loss. As for the unsupervised loss, it is divided into two parts: the loss for image-level consistency $\mathcal{L}_s$ and the loss for feature-level consistency $\mathcal{L}_{fp}$.

We define $x^w\in \mathbb{R}^{C\times H\times R}$ as the weakly perturbed unlabeled training data generated.$x^{s_1}\in \mathbb{R}^{C\times H\times W},x^{s_2}\in \mathbb{R}^{C\times H\times W}$ represent the two sets of strongly perturbed unlabeled training data generated. $p^w\in \mathbb{R}^{H\times W},p^{s_1}\in \mathbb{R}^{H\times W},p^{s_2}\in \mathbb{R}^{H\times W}$ denote the model predictions for the unlabeled training data with different levels of perturbation, respectively. We define the image consistency loss function $\mathcal{L}_s$ as follows:
\begin{equation}
\mathcal{L}_s = \frac{1}{B_u} \sum \mathbbm{1}\left(\max(p^w) \geq \tau\right) \cdot \left(\mathrm{H}(p^w, p^{s_1}) + \mathrm{H}(p^w, p^{s_2})\right)
\end{equation}
In the above equation, $B_u$ represents the batch size of the unlabeled data, $\tau$ is a pre-set confidence threshold used to mitigate the impact of noise on model training, and $\mathrm{H}$ calculates the entropy distance between two sets of predictions.

For the loss of feature-level consistency between strong and weak perturbations $\mathcal{L}_{fp}$, we define $p^{fsp}\in \mathbb{R}^{H\times W},p^{fwp}\in \mathbb{R}^{H\times W}$ to represent the model's predictive outcomes after applying strong and weak perturbations to the features, respectively. We define the loss function for feature consistency $\mathcal{L}_{fp}$ as follows:
\begin{equation}
\mathcal{L}_{fp} = \frac{1}{B_u} \sum \mathbbm{1}\left(\max(p^{fwp}) \geq \tau\right) \cdot \mathrm{H}(p^{fwp}, p^{fsp}) 
\end{equation}
Hereby, we can articulate the expression for the unsupervised loss function as follows:
\begin{equation}\label{eq14}
\resizebox{\linewidth}{!}{$
\begin{aligned}
\mathcal{L}_u = \frac{1}{B_u} \sum \mathbbm{1}\left(\max(p^w) \geq \tau\right)\cdot \left(\frac{\lambda}{2} \left(\mathrm{H}(p^w, p^{s_1})\right) + \mathrm{H}(p^w, p^{s_2})\right) \\
+ \mathbbm{1}\left(\max(p^{fwp}) \geq \tau\right) \cdot \left(\mu \mathrm{H}(p^{fwp}, p^{fsp})\right)
\end{aligned}
$}
\end{equation}
In the unsupervised loss $L_u$, the function $\mathrm{H}$ is commonly used as the Dice coefficient loss.
\section{Experiments}
\begin{figure}
\centering
\includegraphics[width=1.0\linewidth, keepaspectratio]{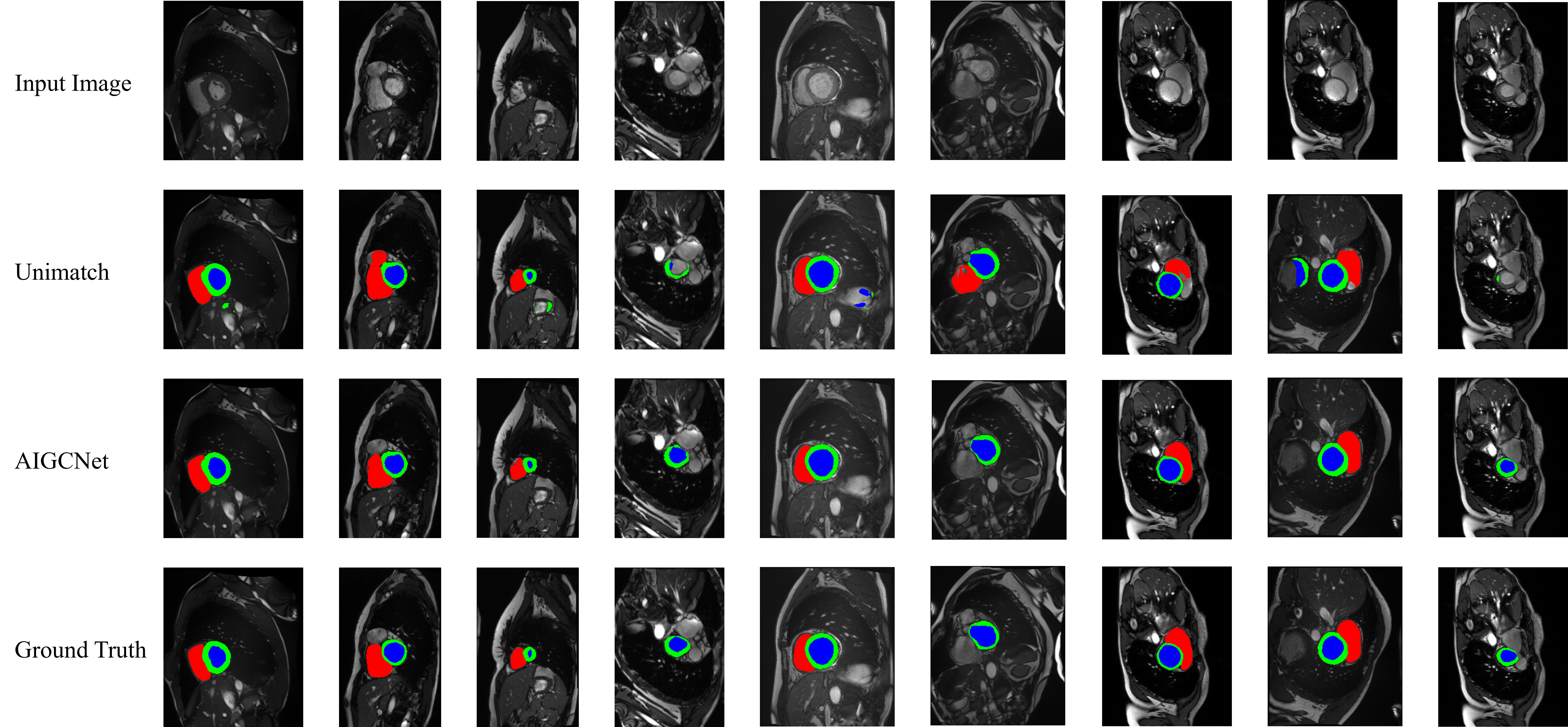}
\setlength{\belowcaptionskip}{-0.4cm} %调整图片标题与下文距离
\caption{Comparison with ground truth and two different results. The three different colors depicted in the figure represent three distinct segmentation outcomes. The section enclosed in a light blue box indicates areas where the model has either made an erroneous prediction or omitted a prediction.}
%\vspace{-1cm}
\end{figure}
\begin{table*}[!t]
\centering
\setlength{\tabcolsep}{1pt}
\resizebox{0.9\textwidth}{!}{
\begin{tabular}{l|c|cccc|cccc|cccc}
\hline
\multirow[c]{2}{*}{Method} & \multirow[c]{2}{*}{Reference} & \multicolumn{4}{c|}{1 case} & \multicolumn{4}{c|}{3 cases} & \multicolumn{4}{c}{7 cases} \\
\cmidrule{3-14}
& & DICE$\uparrow$ & Jacc$\uparrow$ & 95HD$\downarrow$ & ASD$\downarrow$ & DICE$\uparrow$ & Jacc$\uparrow$ & 95HD$\downarrow$ & ASD$\downarrow$ & DICE$\uparrow$ & Jacc$\uparrow$ & 95HD$\downarrow$ & ASD$\downarrow$ \\
\hline
SupBaseline & -- & 28.50 & 16.62 & -- & -- & 47.83 & 37.01 & 31.16 & 12.62 & 79.41 & 68.11 & 9.35 & 2.70 \\

UA\mbox{-}MT~\cite{yu2019uncertainty} & MICCAI’19 & 42.28 & 32.21 & 40.74 & 18.58 & 72.71 & 60.89 & 21.48 & 7.15 & 85.14 & 75.90 & 6.25 & 1.80 \\

ICT~\cite{verma2019interpolation} & IJCAI’19 & 43.03 & 30.58 & 34.92 & 15.23 & 74.20 & 62.90 & 17.01 & 4.32 & 85.15 & 76.05 & 4.27 & 1.46 \\

SASSNet~\cite{li2020shape} & MICCAI’20 & -- & -- & -- & -- & 57.77 & 46.14 & 20.05 & 6.06 & 84.50 & 74.34 & 5.42 & 1.86 \\

FixMatch~\cite{sohn2020fixmatch} & NeurIPS’20 & 69.67 & 58.34 & 37.92 & 14.41 & 83.12 & 73.59 & 9.86 & 2.61 & 88.31 & 79.97 & 7.35 & 1.79 \\

DTC~\cite{luo2021semi} & AAAI’21 & -- & -- & -- & -- & 56.90 & 45.67 & 23.36 & 7.39 & 84.29 & 73.92 & 12.81 & 4.01 \\

CPS~\cite{chen2021semi} & CVPR’21 & 56.70 & 44.31 & 24.97 & 10.48 & 75.24 & 64.67 & 10.93 & 2.98 & 84.63 & 75.20 & 7.57 & 2.27 \\

MCNetV2~\cite{wu2022mutual} & MIA’22 & 57.49 & 43.29 & 31.31 & 10.97 & 78.96 & 68.15 & 12.13 & 3.91 & 85.97 & 77.21 & 7.55 & 2.11 \\

MC\mbox{-}Net~\cite{wu2021semi} & MICCAI’21 & -- & -- & -- & -- & 62.85 & 52.29 & 7.62 & 2.33 & 86.44 & 77.04 & 5.50 & 1.84 \\

URPC~\cite{luo2022semi} & MedIA’22 & -- & -- & -- & -- & 55.87 & 44.64 & 13.60 & 3.74 & 83.10 & 72.41 & 4.84 & 1.53 \\

SS\mbox{-}Net~\cite{wu2022exploring} & MICCAI’22 & -- & -- & -- & -- & 65.82 & 55.38 & 6.67 & 2.28 & 86.78 & 77.67 & 6.07 & 1.40 \\

BCP~\cite{bai2023bidirectional} & CVPR’23 & -- & -- & -- & -- & 87.59 & 78.67 & 1.90 & 0.67 & 88.84 & 80.62 & 3.98 & 1.17 \\

DMD~\cite{xie2023deep} & MICCAI’23 & -- & -- & -- & -- & 80.60 & 69.08 & 5.96 & 1.90 & 87.52 & 78.62 & 4.81 & 1.60 \\

INCL~\cite{zhu2024inherent} & MIDL’23 & 77.80 & 66.13 & 11.69 & 3.22 & 85.43 & 75.76 & 6.37 & 1.37 & 88.28 & 80.09 & 1.67 & 0.49 \\

Unimatch~\cite{Yang_2023_CVPR} & CVPR’23 
&85.43&74.57& -- & -- & 88.86& 79.95 & -- & --  & 89.85& 81.57 & -- & -- \\

%& -- & -- & -- & -- & 84.38 & 75.54 & 5.06 & 1.04 & 88.08 & 80.10 & 2.09 & 0.45 \\

URCA~\cite{qin2024urca} & CMPB’24 & -- & -- & -- & -- & 83.31 & -- & 6.95 & 2.16 & 87.86 & -- & 4.21 & 1.36 \\

CrossMatch~\cite{zhao2024crossmatchenhancesemisupervisedmedical} & arXiv’24 & -- & -- & -- & -- & 88.27 & 80.17 & 1.53 & 0.46 & 89.08 & 81.44 & 1.52 & 0.52 \\

DiffRect~\cite{liu2024diffrect} & MICCAI’24 & 82.40 & 71.96 & 10.04 & 2.90 & 86.95 & 78.08 & 4.07 & 1.23 & 90.18 & 82.12 & 1.38 & 0.48 \\

\textbf{Ours} & -- & \textbf{86.25}$^{\pm2.50}$ & \textbf{75.82} & \textbf{1.11} & \textbf{0.21} & \textbf{89.53}$^{\pm2.15}$ & \textbf{81.07} & \textbf{0.94} & \textbf{0.17} & \textbf{90.40}$^{\pm1.68}$ & \textbf{82.48} & \textbf{0.86} & \textbf{0.14} \\
\hline
\end{tabular}}
\caption{Comparison with different models on the ACDC dataset, using 1, 3, and 7 labeled cases. Methods are ordered chronologically by publication year.}
\end{table*}

\subsection{Experimental Setups}
\subsubsection{\textbf{Datasets}}
In this research, we evaluated our model's performance on the ACDC within a semi-supervised learning framework for medical image segmentation. The ACDC dataset\cite{8360453}, consisting of 100 two-dimensional cine cardiac MR images from the University Hospital of Dijon, is divided into 70 scans for training, 10 for validation, and 20 for testing. It is unique in targeting multiclass segmentation of cardiac structures, contrasting with other datasets that focus on binary segmentation. Within the domain of semi-supervised medical image segmentation, the ACDC dataset is conventionally divided into distinct segments categorized as 1-case, 3-case, and 7-case. These categories correspond to the utilization of 1\%, 5\%, and 10\% of the available labeled data, respectively.

%For the ISIC-2017 dataset, we adopted a similar semi-supervised learning strategy. This dataset, which encompasses a wide range of skin lesion images, is an excellent resource for evaluating segmentation models in the context of dermatology. We utilized 10\% of the labeled images for training, which amounts to 200 images, and the remaining 90\% as unlabeled data for our semi-supervised model. The validation set within ISIC-2017 contains 150 images, which were used to evaluate the model's performance in a real-world scenario.
%For the ISIC-2017 dataset, we adopted a similar semi-supervised learning strategy. This dataset, which encompasses a wide range of skin lesion images, is an excellent resource for evaluating segmentation models in the context of dermatology. We utilized 10\% of the labeled images for training, which amounts to 200 images, and the remaining 90\% as unlabeled data for our semi-supervised model. This setup allowed us to evaluate how well our model could generalize and perform segmentation on unseen data with minimal supervision.
%The ISIC-2017 is a dataset of xxx.it contain 2000 images in xxxx.We adopt 10\% label image and other 90\% as the unlabeled data to evaluate the model's performance in medical image semi-supervised learning.

\subsubsection{\textbf{Implementation Details}}
To ensure a fair comparison with prior work, we just adopted a simple U-Net architecture as the backbone for our segmentation model. Following previous work \cite{Yang_2023_CVPR,Bai_2023_CVPR}, we employ the Dice Similarity Coefficient, the Jaccard Index, the 95th percentile Hausdorff Distance (95HD), and the Average Surface Distance (ASD) to evaluate the segmentation accuracy of our medical image analysis model. We initialized the learning rate at 0.01 and used an SGD optimizer. The model was trained for 300 epochs under a learning rate scheduler. The original images were resized to vary between 0.5 and 2.0 times their original dimensions, and subjected to cropping and flipping to obtain their weakly perturbed versions. The weights for the loss function, denoted as $\lambda$ and $\mu$, are set to 0.5. 
\subsection{Semi-supervised  image segmentation in ACDC}

In the ACDC dataset, we compared our experimental results with those of multiple models, consistently achieving superior results. In comparison to the previous state-of-the-art (SOTA), our method demonstrated improvements across all four metrics.

It should be noted that, compared to previous methodologies, our approach has yielded outcomes equivalent or superior to those of other models that have used 10\% labeled data, although it only employs 5\% labeled data. Furthermore, compared to previous methods, DiffRect, which labeled 5\% labeled data, our approach has achieved comparable outcomes with only 1\% of the data, which has often been overlooked due to its poor performance in previous studies. This substantiates the superior performance and substantial potential of AIGCMatch in the domain of medical image segmentation.

\subsection{Ablation study}
To rigorously assess the effectiveness of each component within our framework, we have meticulously designed and executed a series of systematic experiments. These experiments isolate and evaluate the impact of individual elements within the framework.

\subsubsection{Ablation study on Perturbation at the image and feature level}

Primarily, we aim to establish the validity of the attention-guided perturbation at both the image and feature levels. To achieve this, we introduce the attention-guided perturbation at various levels of the model and compare its performance with the introduction of random noise, a technique employed by UniMatch.

\begin{table}[ht]
\centering
\setlength{\belowcaptionskip}{-0.4cm}
\begin{tabular}{cc|c}
\hline
 Image perturbation  & Feature perturbation  & MeanDice (\%) \\
\hline
Random & Random & 89.85 \\
Random & AttFeaPerb & 90.20 \\
AttCutMix & Random & 89.92 \\
%\hline
\textbf{AttCutMix} & \textbf{AttFeaPerb} & \textbf{90.40} \\
\hline
\end{tabular}
\caption{Comparison with 4 different methods to add perturbation. Random represents adding perturbation randomly while AttPerb represents the attention-guided perturbation method.}
\end{table}
Based on the findings presented in Table 3, it can be observed that the utilization of attention-guided perturbation strategies leads to enhancements in model performance, irrespective of whether the perturbation is applied simultaneously at the image level, the feature level, or across both levels.

\subsubsection{Ablation study on Perturbation methods}

To substantiate the efficacy of AttCutMix and AttFeaPerb, we conducted 4 sets of experiments on the ACDC dataset under three distinct partition schemes: 1 case, 3 cases, and 7 cases. The methodologies used in these experiments were the Baseline, the AttCutMix strategy, the AttFeaPerb strategy, and AIGCMatch, respectively.

\begin{table}[ht]
\centering
\setlength{\belowcaptionskip}{-0.4cm}
\begin{tabular}{c|ccc}
\hline
Method & 1 case & 3 cases & 7 cases \\
\hline
Baseline & 85.43 & 88.86 & 89.85 \\
AttCutMix & 86.56 & 89.13 & 89.92 \\
AttFeaPerb & 85.23 & 89.25 & 90.20 \\
\textbf{AIGCMatch} & \textbf{86.25} & \textbf{89.53} & \textbf{90.40} \\
\hline
\end{tabular}
\caption{Comparison with 4 different methodologies on the ACDC dataset, using 1, 3, and 7 labeled cases.}
\vspace{-0.5cm}
\end{table}

Examination of Table 4 reveals that AttCutMix and AttFeaPerb have improved over the original model across experiments with varying proportions of labeled data. Furthermore, integrating these two methods, AIGCMatch has resulted in state-of-the-art performance.

\section{Conclusions}
In this study, we introduced an easy yet efficient semi-supervised segmentation framework termed Attention-Guided weak-to-strong Consistency Match (AIGCMatch), aiming at enhancing the performance of medical image segmentation. By employing attention-guided perturbation strategies at both the image and feature levels, AIGCMatch achieves weak-to-strong consistency and preserves the structural information of medical images as well. Experiments on the ACDC dataset demonstrate the effectiveness of our approach, and we achieve the performance of SOTA on the ACDC dataset. 
%Our work demonstrates the efficacy of attention-guided perturbation strategies in medical image segmentation, improving model robustness by introducing structured and semantically relevant perturbations within the model's decision-making process.

%However, there is still room for further improvement and expansion. 

Future work will explore the application of AIGCMatch to a broader range of medical image datasets and investigate its performance on more complex and diverse medical images.
\bibliographystyle{ieeetr}
\bibliography{bibm}

\end{document}